# The digital harms of smart home devices: A systematic literature review

David Buil-Gil[1], Steven Kemp[2], Stefanie Kuenzel[3], Lynne Coventry[4], Sameh Zakhary[5], Daniel Tilley[6] and James Nicholson[7]

[1]Department of Criminology, University of Manchester, UK
[2]Department of Law, Pompeu Fabra University, Spain
[3]Department of Electronic Engineering, Royal Holloway University of London, UK
[4]Abertay University, UK
[5]Independent researcher
[6]Daniel Tilley Analytic Solutions Limited, UK
[7]Department of Computer and Information Sciences, Northumbria University, UK

## Abstract

The connection of home electronic devices to the internet allows remote control of physical devices and involves the collection of large volumes of data. With the increase in the uptake of Internet-of-Things home devices, it becomes critical to understand the digital harms of smart homes. We present a systematic literature review on the security and privacy harms of smart homes. PRISMA methodology is used to systematically review 63 studies published between January 2011 and October 2021; and a review of known cases is undertaken to illustrate the literature review findings with real-world scenarios. Published literature identifies that smart homes may pose threats to confidentiality (unwanted release of information), authentication (sensing information being falsified) and unauthorised access to system controls. Most existing studies focus on privacy intrusions as a prevalent form of harm against smart homes. Other types of harms that are less common in the literature include hacking, malware and DoS attacks. Digital harms, and data associated with these harms, may vary extensively across smart devices. Most studies propose technical measures to mitigate digital harms, while fewer consider social prevention mechanisms. We also identify salient gaps in research, and argue that these should be addressed in future cross-disciplinary research initiatives.

## Corresponding author

David Buil-Gil, 2.17 Williamson Building, Department of Criminology, University of Manchester, Oxford Road, Manchester M13 9PL, UK. Email: david.builgil@manchester.ac.uk

## Acknowledgment

This work is funded by the PETRAS National Centre of Excellence for IoT Systems Cybersecurity.

# 1. Introduction

The connection of electronic devices to the internet allows remote control of physical devices and involves remote collection and sharing of large volumes of data. "Internet-of-Things" (IoT) is the term used to refer to physical objects embedded with sensors and software that connect them to other devices and systems over the internet (Atzori et al., 2010; Weber, 2010). Since the early 1980s, when a group of researchers from Carnegie Melon University connected a Coca-Cola vending machine to the internet for the first time, the IoT paradigm has expanded to encompass many different types of physical devices, including corporate security systems, smartphones, connected cars, electrical grids, military equipment, and home appliances. The connection of home electronic devices and attributes to the internet is known as a "smart home" (Lutolf, 1992). Smart homes may incorporate remote-controlled lighting, heating and water consumption, smart meters and internet-connected home security systems, as well as other home devices such as televisions, door locks, remote baby or pet control systems, refrigerators, or voice control devices (e.g., Google Home, Amazon Alexa). Almost any home electrical appliance can be connected to the internet, and multiple interconnected devices form smart home ecosystems. Smart home technologies are used not only to activate and deactivate appliances, but also to monitor the activities of households and automate certain aspects of everyday life (Ricquebourg et al., 2006). The use of IoT home devices is increasingly widespread: in the UK, a survey conducted by techUK and GfK in 2021 showed that 58% of respondents owned a smart TV, 39% smart speakers, 24% smart fitness and activity trackers and 15% smart thermostats (techUK, 2021). In March 2022, 51% of all meter readers in the UK were smart or advanced meters (BEIS, 2022).

While smart homes present many opportunities for users and may improve energy efficiency (Corbett, 2013), they also pose challenges to the security and privacy of users (Ali et al., 2017; Komninos et al., 2014). With the increase in the uptake of IoT home devices, it becomes critical to understand the digital harms that can be associated with smart homes. The main challenge of smart homes is related to the large amount of security-critical and privacy-sensitive data they record from users (Dorri et al., 2017). Lin and Bergmann (2016) argue that smart homes may pose threats to *confidentiality* (i.e., unwanted release of sensitive information), *authentication* (i.e., control or sensing information being falsified) and *access* (i.e., unauthorised access to system controls). For instance, *confidentiality* breaches may lead to an unwanted release of information about electricity usage that may inform potential offenders about the times when a house is not occupied (Blythe and Johnson, 2021; Hodges, 2021). *Confidentiality* breaches may also lead to a loss of sensitive medical data or other types of sensitive information, that can be used by offenders to hold data to ransom (Tzezana, 2016). An *authentication* threat may exist, for example, if an automated fire control system is tampered with to activate the emergency alarm system and unlock all doors, thus enabling anyone to access the building (Jacobsson et al., 2016). Unauthorised *access* to smart home control systems may enable the activation of webcams and voice control devices, or control of home appliances such as ovens or electric stoves, thus making the entire smart home ecosystem insecure. Smart homes may also enable new forms of cyberstalking and exacerbate power asymmetries between household members (Nicholls et al., 2020).

For all these reasons, it becomes crucial to fully understand the digital harms of smart homes, in terms of threats to privacy and security. A growing body of research has begun to speculate about the security and privacy challenges of IoT home devices, and record data about digital harms known to public authorities and users' perceptions and experiences. The field is now at a point where these unique studies can be synthesised to create a comprehensive review of the digital harms of smart homes, which may serve to further inform policy and sociotechnical solutions to mitigate them. This article presents a systematic review of the literature using observational, experimental, documental, or case study research methods to analyse the security and privacy harms of smart home applications and technologies. Previously, Marikyan et al. (2019) conducted a systematic review of studies published between 2002 and 2017 to explore the definitions, services and functions of smart homes and the main motivations for smart home adoption. They highlighted that one of the main barriers to the adoption of smart devices was the high perceptions of privacy and security risks among users. Blythe and Johnson (2021) conducted a systematic review of articles published between 2007 and 2017 to analyse crimes facilitated by consumer IoT. Other systematic reviews have also synthesised the literature about the security challenges of smart cities (Laufs et al., 2020) and applications of smart homes to monitor the well-being of older adults (Demiris and Hensel, 2008). Our research builds on and expands previous literature reviews about the privacy and security harms of smart home devices. More specifically, the aims of this article are:

- Classify the digital harms of smart homes;
- Identify smart home devices and attributes that pose digital harms; and
- Explore policy and sociotechnical approaches to mitigate the digital harms of smart homes.

Thus, to our knowledge, this is the first systematic review of the literature to specifically focus on the digital harms of smart homes. Importantly, the use of smart home appliances has increased rapidly since the last systematic review of crimes facilitated by consumer IoT, which was conducted in 2017 (Blythe and Johnson, 2021), and many new digital harms may have emerged since then. For instance, according to estimates by techUK (2021), the ownership of smart speakers increased by 81% between 2017 and 2021, and this increase was larger than 75% in the case of smart doorbells, 50% in smart lighting, 49% in smart TVs and 47% in smart thermostats.

This article is structured as follows: Section 2 presents an overarching description of recent developments in smart home ecosystems. Section 3 describes the methodology of the systematic review, including the search strategy, selection of studies, and data extraction. Section 4 presents the results, and Section 5 presents the discussion and final conclusions.

## 2. Smart home: Opportunities and digital harms

Marikyan et al. (2019) identified four broad areas in which smart home devices can provide benefits for users: health-related benefits (e.g., detection of dangerous events), environmental benefits (e.g., reduction in electricity consumption), financial benefits (e.g., cheaper virtual visits), and psychological wellbeing and social inclusion (e.g., virtual interaction and entertainment). These benefits coincide with the most relevant benefits found by Sovacool and Furszyfer Del Rio (2021) in their study using expert interviews, though these authors also highlight the relevance of "convenience and controllability" provided by smart homes (see also Lee et al., 2017). However, in addition to the potential benefits, it is also key to understand the risks and barriers of smart home technologies.

There are multiple ways consumer IoT can be exploited for crime (Blythe and Johnson, 2021), and it is easy to find examples of attacks involving smart home devices. Possibly the most famous of these is the Mirai botnet, which exploits poor security in IoT devices and has been used in numerous disruptive DDoS attacks around the world (Krebs, 2017). According to Brian Krebs, the authors of the initial attacks in 2016 published the source code for Mirai, meaning it was reused and sold as a DDoS-for-hire service. In 2017, the developers of the Mirai malware were also found guilty of infecting IoT devices and home routers to create another botnet that was used in a click-fraud scam to generate illicit advertising revenue (US Department of Justice, 2017). Other well-known attacks have involved the hacking of home cameras that are used for security and baby monitoring, thereby allowing private videos to be freely viewed online (BBC, 2013). Relatedly, research has highlighted the role of smart home devices in domestic abuse (Nicholls et al., 2020).

Outlining just a few examples of attacks that have used smart home devices provides an insight into the wide range of potential harms from these technologies. This has not gone unnoticed by government agencies. For instance, in the UK, the Product Security and Telecommunications Infrastructure (PSTI) Bill is currently being processed by the legislator, with the department behind the bill stating that its objective is to protect against "the harms enabled through insecure consumer connectable products" (DCMS, 2021a). This legislation links closely to the concept of safety by design that was explicitly noted in the UK Government Online Harms White Paper from 2019 (DCMS, 2019) and to the definition of "online harms" in the government's draft Online Safety Bill: "user generated content or behaviour that is illegal or could cause significant physical or psychological harm to a person" (DCMS, 2021b).

However, this official definition may not cover the first two examples of botnet-based attacks described above, since these do not necessarily cause physical or psychological harm to a person. Thus, to fully understand and analyse the digital harms related to smart home devices a more tailored definition and classification is necessary. Unfortunately, despite the clear policy interest in harms from smart home devices (Piasecki et al., 2021), an agreed-upon taxonomy does not exist. This is problematic because to prevent digital harms we first need to understand how these might arise. Establishing and prioritising policy responses necessitate a comprehensive assessment of potential harms (Agrafiotis et al., 2018).

To this end, we have adapted classifications of online harms by McGuire and Dowling (2013), Wall (2001), and Lin and Bergmann (2016) in accordance with the nature, objective, and method of online harm, respectively (Table 1). Firstly, with regard to the nature of harm, this is divided into cyber-dependent harms that can only occur online, such as DDoS attacks, and cyber-enabled harms that can also take place offline but are increased in scope by the internet, for example, fraud or stalking. In the second place, the objective of the harm is more akin to a legal categorisation. It includes (a) 'cyber-trespass' when invisible boundaries are crossed, such as hacking a computer system, (b) 'cyber-deception and theft', for example, the myriad of possible frauds committed over the internet, (c) 'cyber-porn and obscenity', which can sometimes not necessarily be illegal, and (d) 'cyber-violence', which involve injurious or hurtful behaviour such as stalking. Finally, we adapt a classification of the method used to bring about the harm (Lin and Bergmann, 2016). This can be achieved by the unwanted release of information (confidentiality), falsification of control or sensing information (authentication), or unauthorised access to system controls (access). We will apply this classification to record information about digital harms from articles included in the systematic literature review.

*Table 1. Proposed classifications of online harms*

| Classification | Definition | Examples |
| --- | --- | --- |
| ***According to nature of harm.*** *Adaptation of classification by McGuire and Dowling (2013)* | | |
| Cyber-dependent harm | Harms that can only be committed through the internet and do not have an equivalent offline mode | Malware, DoS, hacking |
| Cyber-enabled harm | Harms that have an offline equivalent mode but have increased in reach and impact due to the internet | Fraud, stalking, grooming |
| ***According to objective of harm.*** *Adaptation of classification by Wall (2001)* | | |
| Cyber-trespass | Crossing of invisible boundaries of ownership online | Hacking, access to private/confidential data |
| Cyber-deception and theft | Harmful or criminal acquisitions that occur online | Fraud, identity theft, digital piracy |
| Cyber-porn and obscenity | Deviant content related with sex and pornography | Pornography, sexual services, online child sexual exploitation |
| Cyber-violence | Injurious, hurtful or dangerous online materials | Stalking, harassment, terrorism |
| ***According to method of harm.*** *Adaptation of classification by Lin and Bergmann (2016)* | | |
| Confidentiality | Unwanted release of sensitive information | Release of information about electricity usage, explicit photos |
| Authentication | Control or sensing information being falsified | False data injection, system tempered with to unlock doors |
| Access | Unauthorised access to system controls | Activation of web cam, control of voice assisted device |

## 3. Methodology

This article takes a two-fold methodological approach to synthesising the recent literature about the digital harms related to smart homes. First, we systematically review all relevant studies published between January 2011 and October 2021. By systematically reviewing the literature we aimed to classify the digital harms related to smart homes, identify the smart home devices and attributes that pose digital harms, and explore potential policy and sociotechnical approaches to mitigate digital harms. We have restricted our search to studies published since 2011 due to the rapid technological development of smart technologies and to facilitate the search. Second, where possible, we illustrate the findings of the literature review with real-world cases.

We conduct the systematic literature review using a-priori criteria to search, select and extract data from studies. The systematic review protocol follows the Preferred Reporting Items for Systematic reviews and Meta-Analyses for Protocols 2015 (PRISMA), which is a widely used checklist to facilitate the design of robust protocols for systematic reviews (Mohler et al., 2015). The following sections explain the systematic review protocol in detail.

3.1 Search strategy

We have selected those articles that use observational, experimental, documental, or case study research methods to analyse the security and privacy harms of smart homes. Thus, we do not include theoretical or technical notes or reviews of the literature. We have included peer-reviewed studies published in English. Since this article is particularly interested in smart homes, we have excluded all studies that analyse related technologies in settings that are not solely residential (e.g., IoT for cities, business, healthcare or any other context). We have also excluded those studies that explore smart homes but do not consider their digital harms, either privacy- or security-related.

The search for published studies was conducted in October 2021. The following databases were used to search for published articles: Web of Science and Scopus. Both databases provide access to multiple multidisciplinary and regional citation indices. Web of Science covers more than 182 million records in engineering, social sciences, natural sciences, biomedical sciences and arts and humanities, with its strongest coverage in engineering, computer science, natural sciences and material sciences. It covers several databases such as the Web of Science Core Collections, BIOSIS, SciELO and Data Citation Index. Scopus includes more than 77 million items from more than 5,000 publishers in many different fields, including computing, information sciences, law, human society, engineering and architecture. Major publishers included in the Web of Science database include Springer, Nature, Wiley, IEEE, Elsevier and ACM. Thus, Web of Science and Scopus include many other digital libraries, such as IEEE Xplore and ACM Digital Library.

The search strategy used the following search terms in titles, abstracts, keywords and subject headings:

> (((SMART) OR (IOT) OR (INTERNET OF THINGS) OR (AUTOMAT) OR (VIRTUAL)) AND ((HOME) OR (HOUSE) OR (DOMESTIC) OR (RESIDEN)) OR (DOMOTICS)) AND ((SECUR) OR (PRIVA) OR (CRIM) OR (HACK) OR (ATTACK) OR (INCIDENT) OR (BREACH) OR (LEAK) OR (HARM) OR (THEFT))

Some terms were truncated to include all related terms. For example, "AUTOMAT" includes automation, automated and automating, "HACK" includes hack, hacking and hacker, and "SECUR" includes "secure", "security" and "cybersecurity". The search strategy was agreed among all co-authors after consulting several practitioners working in public and private sector organisations.

## 3.2 Selection of studies

All identified citations were imported into a database. Duplicated citations were removed. Two researchers then screened the titles and abstracts of all articles against our inclusion criteria, namely: (a) main focus is on smart home appliances or attributes, (b) explores privacy and/or security harms, (c) uses data recorded from observation, case studies, documents or experiments, either quantitative or qualitative, and (d) is available in English. In order to ensure consistency among data collectors, we then selected random samples of 100 citations and shared them with five additional researchers, who also screened the titles and articles against our inclusion criteria. Interrater reliability scores were then calculated, showing moderate-strong levels of agreement for criteria (a) (% agreement = 84.4, Cohen's κ = 0. 67, p-value < 0.001) and (b) (% agreement = 83.8, Cohen's κ = 0. 68, p-value < 0.001), and moderate levels of agreement for criteria (c) (% agreement = 80.0, Cohen's κ = 0. 50, p-value < 0.001) and (d) (% agreement = 97.0, Cohen's κ = 0. 53, p-value < 0.001). The interrater reliability was moderate-strong for the overall inclusion of studies (% agreement = 89.0, Cohen's κ = 0. 68, p-value < 0.001). Disagreements were resolved through consensus between the two primary judges. The main reasons for excluding articles were that the primary focus of the study was not on smart homes (e.g., studies on smart cities, smart farms, smart automobiles), did not study harms (e.g., studies on energy efficiency, regulatory requirements, military applications, security perceptions) or did not use data obtained from observational, case, documental or experimental studies (e.g., literature reviews, technical reviews, theoretical pieces). We also excluded 111 studies that either were not available in English or not available at all.

While interrater reliability indices show a moderate-strong degree of inter-judge reliability, the volume of selected studies was still too large for an exhaustive review of studies (k = 625). We thus considered a fifth inclusion criterion that reduced the number of selected studies: (e) analyses of real-world harms on real-world smart home appliances or attributes, thus excluding both laboratory experiments that do not attack real-world devices and computer simulations not based on real-world data. Finally, 67 studies met the inclusion criteria and were subject to an in-depth review (see PRISMA flowchart in Figure 1). After reviewing the content of selected articles, 4 studies were excluded from the analysis due to failing to meet at least one of our main selection criteria (i.e., 2 did not study digital harms, and 2 did not analyse data obtained from observational, case, documental or experimental studies). 63 studies finally were included in the literature review. A short description of each study is included in Table 2, including a unique identification number for each study, which will be used to refer to it in text.

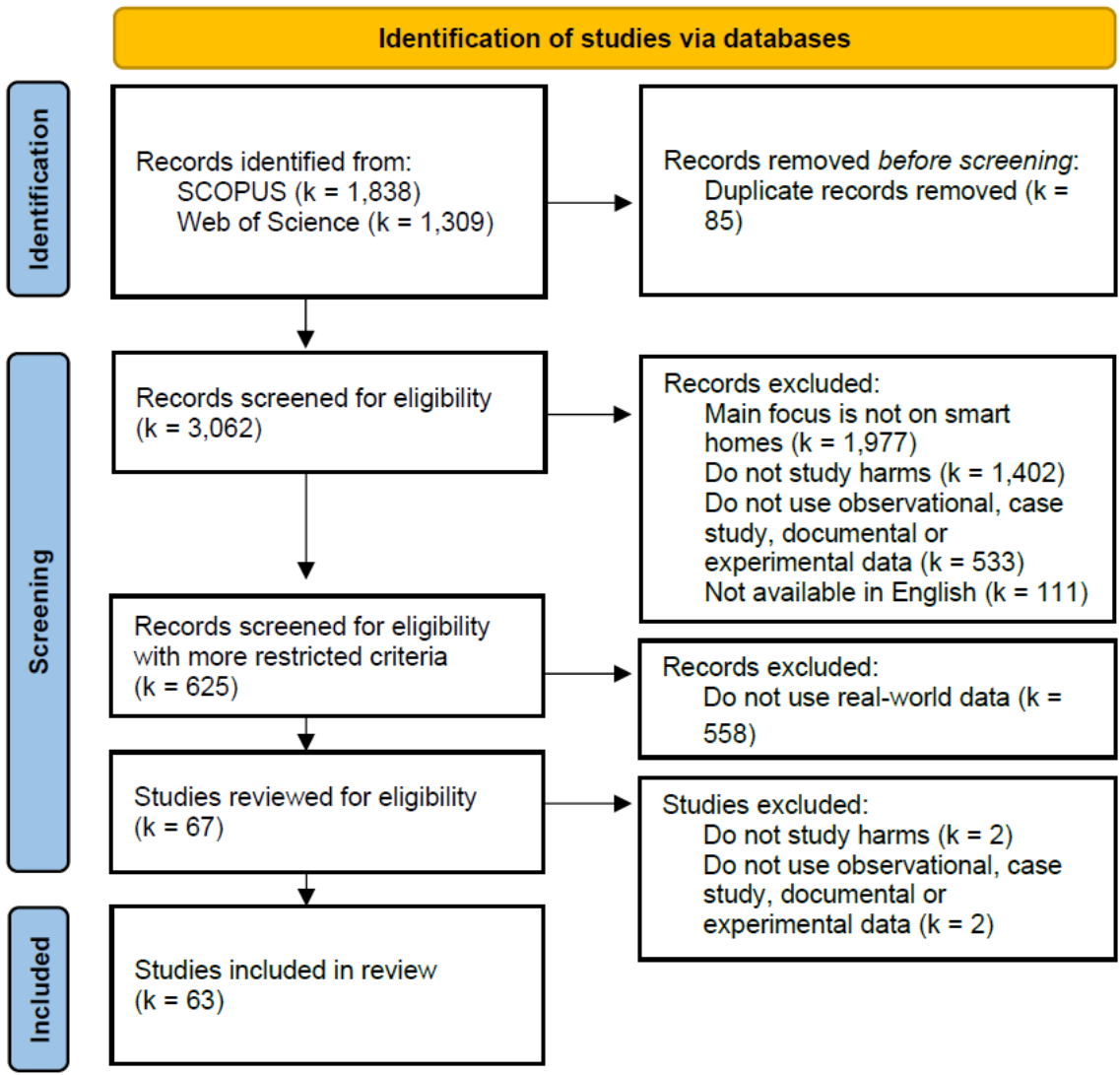

**Figure 1.** PRISMA flowchart in the selection of studies

*Table 2.* Summary description of primary studies included in the systematic review

|    | Paper | Topic of study |
|----|-------|----------------|
| 1  | Javed et al. (2021) | Spoofing countermeasures against voice assistants |
| 2  | Anthi et al. (2021) | Adversarial Machine Learning against Intrusion Detection Systems |
| 3  | OConnor et al. (2021) | Protecting companion apps against man-in-the-middle attacks |
| 4  | Yu et al. (2021) | Sensitive information in the metadata of encrypted packets |
| 5  | Tran et al. (2021) | Voice replay and injection attacks against voice assistants |
| 6  | Wang, Li et al. (2021) | Decision tree models to detect attacks against smart devices |
| 7  | Li et al. (2021) | Automation of privacy policy statements for smart home apps |
| 8  | Tushir et al. (2021) | Impact of DoS attacks on smart devices connected to WiFi |
| 9  | AlOtaibi and Lombardi (2021) | Sound- and network-based attacks against Amazon Echo |
| 10 | Yamauchi et al. (2021) | Machine learning to detect anomalous behaviour |
| 11 | Aafer et al. (2021) | Technical vulnerabilities of Android Smart TV |
| 12 | Wang, Ji et al. (2021) | Automation of security assessment of IoT messaging protocols |
| 13 | Wan et al. (2021) | Unveiling of smart devices from network data |
| 14 | Rauti et al. (2021) | Man-in-the-browser attacks to intercept/modify data |
| 15 | Heartfield et al. (2021) | Self-configurable automated intrusion detection system |
| 16 | Choi et al. (2021) | Older adults' experiences with smart devices |
| 17 | Cultice et al. (2020) | Machine learning to detect anomalous data |
| 18 | Alsheakh and Bhattacharjee (2020) | Automated quantification of security of smart devices |
| 19 | Gassais et al. (2020) | Self-configurable automated intrusion detection system |
| 20 | Peng and Wang (2020) | Network-based monitoring platform to identify security threats |
| 21 | Xiao et al. (2020) | Authentication framework to protect devices from attacks related to open ports and over-privilege |
| 22 | Salomons et al. (2020) | Hardware and control model to protect data in water meters |
| 23 | Wang et al. (2020) | Inferred voice commands against voice assistants |
| 24 | Sikder et al. (2020) | Access control system for multiple users and devices |
| 25 | Li et al. (2020) | Identification of user behaviour from traffic data of cameras |
| 26 | Zainab et al. (2020) | Machine learning to identify spam in smart devices |
| 27 | Bugeja et al. (2020) | Smart devices' software vulnerabilities to DoS attacks |
| 28 | Vidal-González et al. (2002) | Malware attacks against smart homes |
| 29 | Bistarelli et al. (2020) | Malware attacks against smart homes |
| 30 | Hariri et al. (2020) | Man-in-the-middle attack to exploit the heartbeat of devices |
| 31 | Skowron et al. (2020) | Machine Learning to identify devices and users' activities |
| 32 | Javed and Rajabi (2020) | AI-based solution for malicious traffic detection |
| 33 | Sikder et al. (2019) | Markov Chain Machine Learning to detect malicious activity |
| 34 | Leitão (2019) | Smart devices as attack vectors for intimate partner violence |
| 35 | Kennedy et al. (2019) | Voice command fingerprinting attacks against home speakers |
| 36 | Martin et al. (2019) | Malware against Raspberry Pi smart devices |
| 37 | Ullrich et al. (2019) | Vulnerabilities of the firmware of robot vacuum cleaners |
| 38 | Zhang et al. (2019) | Blockchain-based security protocol to protect IoT networks |
| 39 | Alkhatib et al. (2019) | Developers' insights into the privacy of elderly monitoring devices |
| 40 | Mahadewa et al. (2018) | Integrated perspective to identify vulnerabilities of smart homes |
| 41 | Zhang et al. (2018) | Identification of malicious smart home apps |
| 42 | Jia et al. (2018) | Graph-based mechanism to identify vulnerabilities of smart homes |
| 43 | Anthi et al. (2018) | Security of adaptive IoT hub for smart home ecosystems |
| 44 | Isawa et al. (2018) | Disassembly-code-based similarity between IoT malware |
| 45 | Bhatt and Morais (2018) | Anomaly detection system for smart homes |
| 46 | Do et al. (2018) | Adversarial models to identify vulnerabilities of smart devices |
| 47 | Bordel et al. (2018) | Large datasets reduction for smart home security systems |
| 48 | Sivanathan et al. (2018) | Penetration testing to assess the security of consumer IoT devices |
| 49 | Lally and Sgandurra (2018) | Framework to evaluate vulnerabilities of smart devices |
| 50 | Ji et al. (2018) | Eavesdropping of smart wireless cameras |
| 51 | Mashima et al. (2018) | Estimation of sensitive information from energy usage data |
| 52 | Teng et al. (2017) | Over-the-air firmware update system for routers and gateways |

| 53 | Fan et al. (2017) | Obfuscation of reactive power demand of smart meters |
| 54 | Lyu et al. (2017) | The capacity of consumer IoT devices to participate in DDoS attacks |
| 55 | Sivanathan et al. (2017) | Flow-based network monitoring to identify attacks |
| 56 | Han and Park (2017) | Push button configuration to detect unintended paired devices |
| 57 | Capellupo et al. (2017) | Identification of security vulnerabilities of smart devices |
| 58 | Birchley et al. (2017) | Ethical issues of smart health devices |
| 59 | Copos et al. (2016) | Network traffic analysis to infer sensitive household information |
| 60 | Min and Varadharajan (2016) | Feature-distributed malware to compromise internet services of IoT |
| 61 | de Morais et al. (2014) | Active in-database processing to protect sensitive data of AAL |
| 62 | Matern et al. (2013) | Detection of events in AAL systems using sensor data |
| 63 | Boise et al. (2013) | Older adults' experiences with unobtrusive home monitoring |

## 3.3 Data extraction

Two researchers then reviewed all selected articles and extracted data from them using a standardised form. Aside from information about the year of publication, type of publication, authors, and name of the journal or conference proceedings, which was downloaded automatically from the databases, we recorded detailed information from each article and coded the data into the following categories: (a) design of the study, (b) main aims, (c) type of data analysed, (d) research field, (e) country of authors, (f) country and agency that provided funding, (g) country where data was recorded, (h) smart devices analysed, (i) digital harms identified (by harm type, and according to the categorisation presented in Table 1), (j) type of data that pose a threat or vulnerability, (k) policy or sociotechnical recommendation to mitigate harms, (l) focus of recommendation, and (m) other relevant findings.

For each of these variables, we coded articles according to predefined categories and free-text descriptions with detailed information. For example, to code the design of the study, we distinguished between descriptive, correlational, experimental, meta-analysis, and other types of studies, and then coded all details about the design of each research. Similarly, to study the focus of the recommendation, we distinguished between studies that propose harm prevention or reduction measures focused on the perpetrator, target (e.g., smart device, data), user, or guardian (e.g., third parties that can protect the target or user, such as manufacturers that monitor emerging harms, family members) (Leukfeldt and Yar, 2016), as well as details about the specific recommendation proposed. We also counted the number of citations of each article according to Google Scholar on 20 April 2022. We will present descriptive statistics and tables for every variable recorded in order to summarise the existing evidence about the digital harms of smart devices.

## 3.4 Exemplar cases

In order to further illustrate some of the main findings from the systematic literature review, we accessed detailed information from real-world recorded cases and will present some anonymised descriptions in the "Results" section. More specifically, we obtained details from real cases reported to different police forces in the UK, organisational data breaches sentenced in court in the US and recorded in the Privacy Rights Clearinghouse (PRC) website (https://privacyrights.org/data-breaches), anonymised online reports to websites such as BitcoinAbuse (https://www.bitcoinabuse.com/), and media reports. We purposively and non-randomly select cases to illustrate common themes that arise from the systematic literature review. All accounts of real cases will be anonymised and described in general terms to preserve confidentiality.

# 4. Results

In Section 4.1 we present details about selected studies, including the research field and country of authors, organisations that provide funding, design of the study, aims, and type of data analysed. In Section 4.2 we describe the main types of harms identified and classify them. Finally, in Section 4.3 we summarise the approaches recommended to mitigate the digital harms of smart devices.

## 4.1 Description of selected studies

Amongst the 63 selected studies, 21 (33.3%) of them were published in journals and 42 (66.7%) in conference proceedings. The main journals in the sample were IEEE Access (3), Computers and Security (2), IEEE Internet of Things Journal (2) and Sensors (2); while only two conference proceedings were represented more than once (i.e., Proceedings of the 13th ACM Conference on Security and Privacy in Wireless and Mobile Networks, and Proceedings of the 30th USENIX Security Symposium). IEEE was the most frequent publisher both for journal articles and conference proceedings (7 and 22, respectively), followed by Elsevier (4) and MDPI (3) for journal articles, and Springer (8) and ACM (7) for conference proceedings.

As shown in Figure 2, there is an increase in the frequency of selected studies over time, with 2021 and 2020 being the years with the largest number of articles (16). We note however that data was recorded in October 2021, which in turn means 2021 was the year with the largest ratio of articles per month.

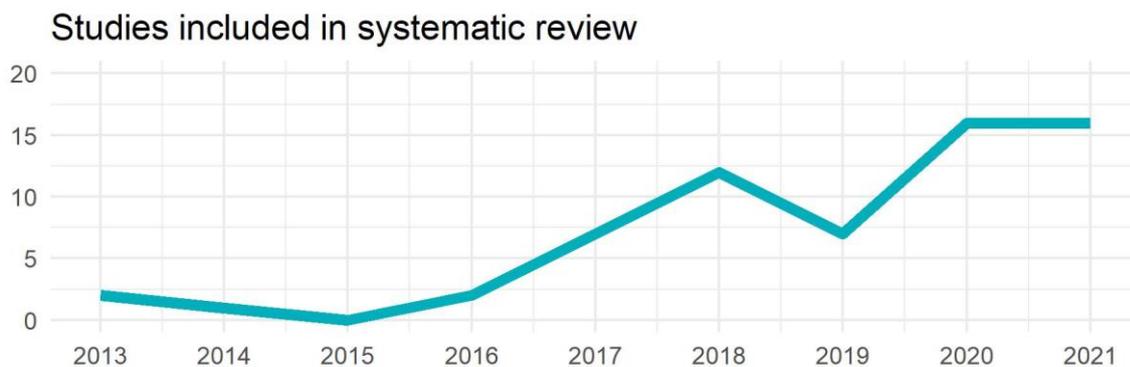

*Figure 2.* Studies included in systematic review by year of publication

While the sample of selected studies included researchers from across 23 countries[1], three countries were represented in the majority of studies: USA (25, 39.7%), China (12, 19.0%) and UK (9, 14.3%), as shown in Figure 3(a). 13 studies involved authors from across multiple countries. Similarly, as shown in Figure 3(b), amongst those studies that acknowledge a source of funding (i.e., 43 out of 63), the main countries (or group of countries) that provide funding for research are USA (17, 39.5%), China (11, 25.6%), European Union (4, 9.3%) and UK (3, 7.0%). The most frequently mentioned funding entities were the USA National Science Foundation (10, 23.3%), Chinese National Natural Science Foundation (5, 11.6%), Chinese National Key Research and Development Program (4, 9.3%), and UK Engineering and Physical Sciences Research Council (3, 7.0%). 24 studies mentioned more than one funding source, and 7 of them obtained funding from more than one country. Aside from national research councils, some studies also acknowledged receiving internal funding from universities, and in some cases from private organisations such as Ericsson, Intel and Schneider. We also recorded information about the places where data were originally recorded, noticing that most of them recorded data in USA (21), UK (9), China (7) and Australia (5). One study analysed data recorded in 5 different countries [28].

---
[1] Studies in the sample included researchers from Australia, Brazil, Canada, China, Finland, Germany, Israel, Italy, Japan, Kazakhstan, Norway, Pakistan, Poland, Portugal, Qatar, Saudi Arabia, South Korea, Singapore, Spain, Sweden, Taiwan, UK and USA.

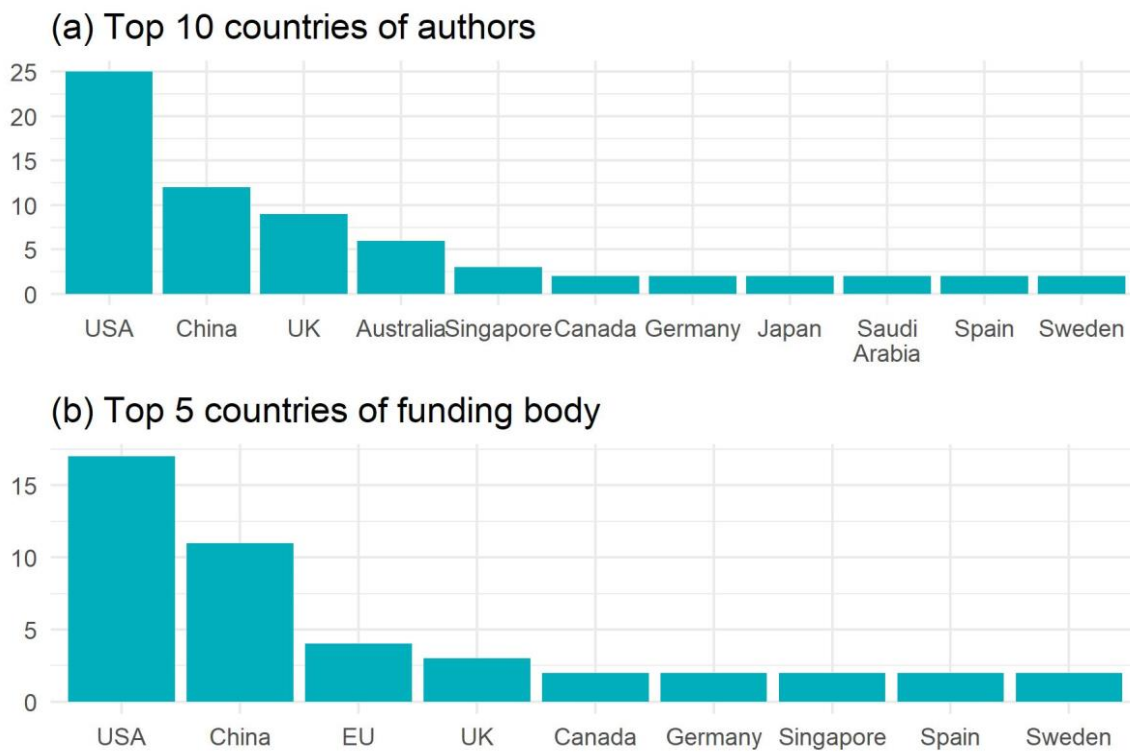

*Figure 3.* Main countries of authors and funding bodies of studies included in systematic review

Regarding the research fields of the authors (compiled from affiliations to university departments and research centres, and authors' bios included in publications), as shown in Figure 4, most authors were affiliated to computer science (39, 61.9%) or computer engineering (28, 44.4%) departments or centres, and fewer to electrical engineering (15, 23.8%). Only 4 studies involved researchers from health disciplines and 1 from social sciences and humanities. We also noted an overall lack of interdisciplinary work, with very few studies involving researchers from different technical disciplines, and not a single study involving researchers from both technical and health or social sciences.

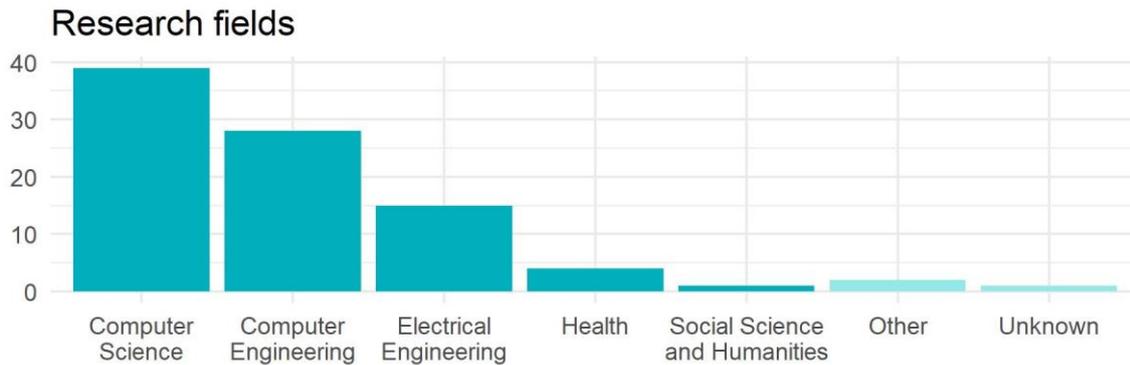

*Figure 4.* Research field of studies included in systematic review

The majority of studies (52, 82.5%) were described as experimental or quasi-experimental by design (i.e., introducing a change, such as exploiting a vulnerability or applying software updates, to experimental groups, such as smart devices or smart home ecosystems, to identify effects in the outcome variables). For instance, [1] tests a spoofing countermeasure for voice assistant systems (e.g., Google Home, Amazon Alexa) against a set of voice spoofing attacks, including synthetic voice attacks and cloned replay attacks; [33] evaluates the effectiveness of a context-aware security framework based on Markov processes to detect malicious actions in smart homes; [48] rates the security level of IoT devices through penetration testing tools; and [50] eavesdrops wireless smart camera traffic to identify the presence of people in the house. Fewer studies were identified as descriptive by design (22, 35.0%). Some examples include [28], [29], and [36], which used honeypots to capture active malware targeting smart homes, thus allowing researchers to examine the characteristics of identified malware; and [39], which used semi-structured interviews with developers to gain a better understanding of privacy issues of aged care monitoring devices. 11 studies combined descriptive and experimental designs. No study followed correlational or meta-analytical designs. The information about the methodological design of studies was recorded from their methods' descriptions, regardless of the overall quality of the design of the study (e.g., sample sizes, randomisation processes, significance tests). We return to this point in the Discussion section.

The majority of studies recorded primary quantitative data (52, 82.5%), such as the volume and characteristics of metadata of encrypted packets sent from smart devices [4], data sent from smart devices to web browser extensions [14], sensor device events [33], or surveys to older adults [63]. Primary qualitative data, including open code of smart apps [7], interviews with engineering researchers [58], and interviews and workshops with survivors of intimate partner violence and support workers [34], was recorded in 15 studies (23.8%). Finally, 8 studies analysed secondary quantitative data, including existing datasets of voice spoofing attacks [1] and real-world cyber-attacks and traffic data [18]; and [34] analysed secondary qualitative data from discussions in domestic abuse forums. 11 studies analysed both quantitative and qualitative data.

With respect to the main objectives of the studies included in our sample, most of them aimed to study the vulnerabilities of specific smart home devices (34, 54.0%), followed by designing and/or evaluating technology solutions to reduce the digital harms of smart homes (33, 52.4%), and studying vulnerabilities of smart homes ecosystems beyond specific devices (27, 42.9%). 27 studies aimed to identify vulnerabilities of smart homes and develop technology solutions. As an example, [20] designed and evaluated a network-based monitoring platform to identify security threats against smart devices, and [23] executed attacks against smart speakers to infer voice commands and then proposed a differential privacy approach to protect such data.

We also recorded data about the number of citations of studies, showing a mean of 18.44 (min = 0, max = 122, median = 8). [41] was the study with the largest number of citations, 122, followed by [59] (114 citations) and [63] (112 citations).

## 4.2 Classifying the digital harms of smart homes

Firstly, we classified the digital harms identified in each study according to the type of incident, including cybercrimes listed by the UK Crime Prosecution Service (n.d.) (e.g., hacking, malware, DoS, stalking) and privacy intrusions more generally. As shown in Figure 5, privacy intrusions were the most common type of harm identified (31, 72.1%), followed by hacking (29, 67.4%), malware (22, 51.2%), DoS/DDoS (21, 48.8%) and stalking (3, 7.0%). Most studies identified different types of harms. Moreover, certain incidents can comprise different harms simultaneously. Examples of these types of harms, both obtained from the systematic literature review and our exemplar cases, are presented in Table 3.

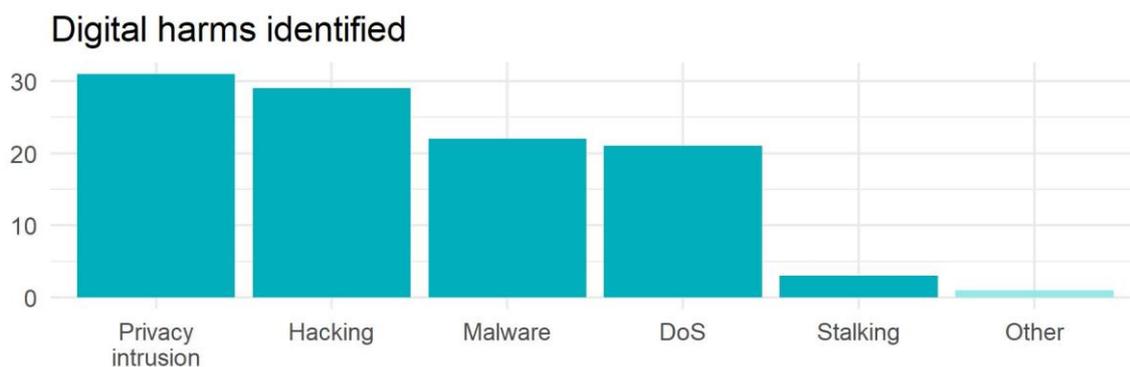

*Figure 5.* *Digital harms (by type of incident) of smart homes*

*Table 3.* Examples of digital harms identified in literature review and case studies

|  | From literature review | From exemplar cases |
|---|---|---|
| Privacy intrusion | Breaches of smart water meters reveal home activities [22] Uncontrolled/unauthorised access to private data recorded by aged care monitoring devices [39] | Smart doorbell camera invades neighbour's privacy (UK court case[2]) App companion of smart sex toy records private moments without consent of the user (USA court case[3]) |
| Hacking | Voice replay and voice injection attacks on voice assistants [5] False data injection on smart devices [26] | Smart TV hacked to access victim's personal details (UK police report) Smart camera and baby monitor feeds from 700 households were hacked and published online (USA court case[4]) |
| Malware | 652,881 interactions with botnets targeting IoT devices [28] 8,713 IoT malware samples [44] | Mirai malware disables CCTV, routers, and other devices (USA court case[5]) Botnet targeting smart home devices and requesting ransom (Bitcoinabuse report) |
| DoS/DDoS | Semantic DoS attacks on five smart home devices [27] DoS attacks on seven routers [52] | Devices infected with Mirai malware to carry our DDoS attacks (USA court case[6]) DDoS attacks on gaming networks (USA[7] and Finland[8] court cases) |
| Stalking | Controlling partner activities through smart cameras, thermostats, TVs and locks [34] Inferring activity of household members from smart thermostat and air detector [59] | Control of ex-partner's activities through Amazon Alexa (UK police report) App companion of ELAN smart home system used to control ex-partner activities (UK police report) |

One study identified a different type of harm that did not fall within the previous categories: traditional access control mechanisms in smart homes consider one unique type of trusted user (in binary terms: control or absence of control), which may lead to certain users being granted an undesired full access control to all devices in the smart home ecosystem [24]. In turn, the authors propose a platform to manage access rights for multiple devices and users.

Secondly, we classify the harms of smart homes according to the taxonomies presented by McGuire and Dowling (2013), Wall (2001), and Lin and Bergmann (2016), which had been previously explained in Section 2. Table 4 summarises the frequencies of studies that identified digital harms according to these three classifications. As before, most studies identified several types of harms and thus are counted in various categories. Based on McGuire and Dowling

---

[2] https://www.judiciary.uk/wp-content/uploads/2021/10/Fairhurst-v-Woodard-Judgment-1.pdf
[3] https://www.courthousenews.com/wp-content/uploads/2018/01/Lovense.pdf
[4] https://www.ftc.gov/system/files/documents/cases/140207trendnetdo.pdf
[5] https://www.justice.gov/usao-nj/press-release/file/1017616/download
[6] https://www.justice.gov/usao-nj/press-release/file/1017616/download
[7] https://www.justice.gov/usao-ndil/file/900826/download
[8] https://www.kaleva.fi/17-vuotias-tuomittiin-murtautumisesta-yli-50-000-p/1842675

(2013), the majority of studies identified cyber-dependent harms (58, 92.1%) – that is, incidents that can only take place online and do not have an equivalent offline mode. Only 8 studies identified cyber-enabled harms. Based on Wall (2001), 59 out of 63 studies (93.7%) focused on cyber-trespass (i.e., crossing online boundaries of ownership), while 9 studies identified harms related to cyber-deception (i.e., harmful acquisitions that occur online, such as identity theft or fraud). Only 3 studies identified cyber-violence and 2 cyber-porn/obscenity. Finally, according to the classification proposed by Lin and Bergmann (2016), 41 studies (65.1%) identified confidentiality, 41 (65.1%) access, and 24 (38.1%) authentication harms.

*Table 4. Classification of digital harms identified in systematic review*

|  |  | Trespass | Deception and theft | Porn and obscenity | Violence |
|---|---|---|---|---|---|
| Cyber-dependent | Confidentiality | 34 | 5 | 1 | 2 |
|  | Authentication | 23 | 6 | 1 | 2 |
|  | Access | 38 | 8 | 1 | 2 |
| Cyber-enabled | Confidentiality | 7 | 0 | 1 | 2 |
|  | Authentication | 3 | 1 | 0 | 1 |
|  | Access | 4 | 1 | 0 | 1 |

As shown in Table 4, most studies focused on harms at the intersection of cyber-dependent, trespass, and access (38, 60.3%). For example, [21] identify harms related to the penetration of smart devices through exploiting open ports and over-privilege of companion apps, and [36] explore malware used to access Raspberry Pi IoT devices with weak credentials. 34 studies (54.0%) focused on harms at the intersection of cyber-dependent, trespass and confidentiality. For example, [59] applies network traffic analysis of data recorded by smart thermostats and air quality detectors to infer sensitive information about events occurring in a property, and [53] analyses reactive power data from smart meters to infer appliance usage information.

These types of harms have also been identified in our exemplar cases. For instance, in 2014, footage from 17 properties in the North East of England was hacked and live-streamed on a Russian website (UK police report), which aligns with the cyber-dependent, trespass, and confidentiality grouping. Similarly, in 2020 there was a Class Action Complaint against Ring LLC in the US arguing that weak software security of smart cameras and doorbells allowed hackers to gain access to the control of these devices[9], which would be an example of cyber-dependent, trespass, and access harm. An example of a cyber-dependent, trespass and authentication harm is identified in [26], which reports false data injections in smart home devices.

While these are the main types of harms identified in the systematic literature review, real-world examples of harms of smart devices exist for all groups in the taxonomy. For instance, a UK celebrity is currently facing trial for posting CCTV feeds of himself having sex with his ex-partner on various porn websites (UK police report), which would fall within a cyber-dependent, porn, and confidentiality group. An example of a cyber-dependent and cyber-enabled, access and violence incident was seen in the hacking of Ring smart cameras and doorbells in 2020, which enabled perpetrators to threaten and racially abuse victims. And an example of cyber-enabled, violence and confidentiality harm can be found in a UK police report of someone who stalked his ex-partner through the app companion of an ELAN smart home system (UK police report).

---
[9] https://www.classaction.org/media/lemay-et-al-v-ring-llc.pdf

### 4.3 Identifying smart home devices that may pose digital harms

We also recorded information about the types of devices associated with digital harms, and visualised results in Figure 6. The most referenced devices in our systematic review were security and surveillance systems (21, 33%), followed by lighting systems and smart bulbs (18, 28.6%) and voice control devices (15, 23.8%). For instance, [25] identifies user behaviour from traffic data of smart cameras, and [42] apply graph-based mechanisms to analyse traffic between Google Home smart speaker and TP-LINK light bulbs and identify vulnerabilities. Before we had seen several exemplar cases of harms related to security systems, such as the UK court case that concluded smart doorbell cameras invade neighbour's privacy, and the USA court case about home CCTV disabled by Mirai malware. Other types of devices that were less commonly referenced included temperature and ventilation devices (12, 19.0%), companion apps and browsers (12, 19.0%), and occupancy-aware control systems (10, 15.9%), amongst others.

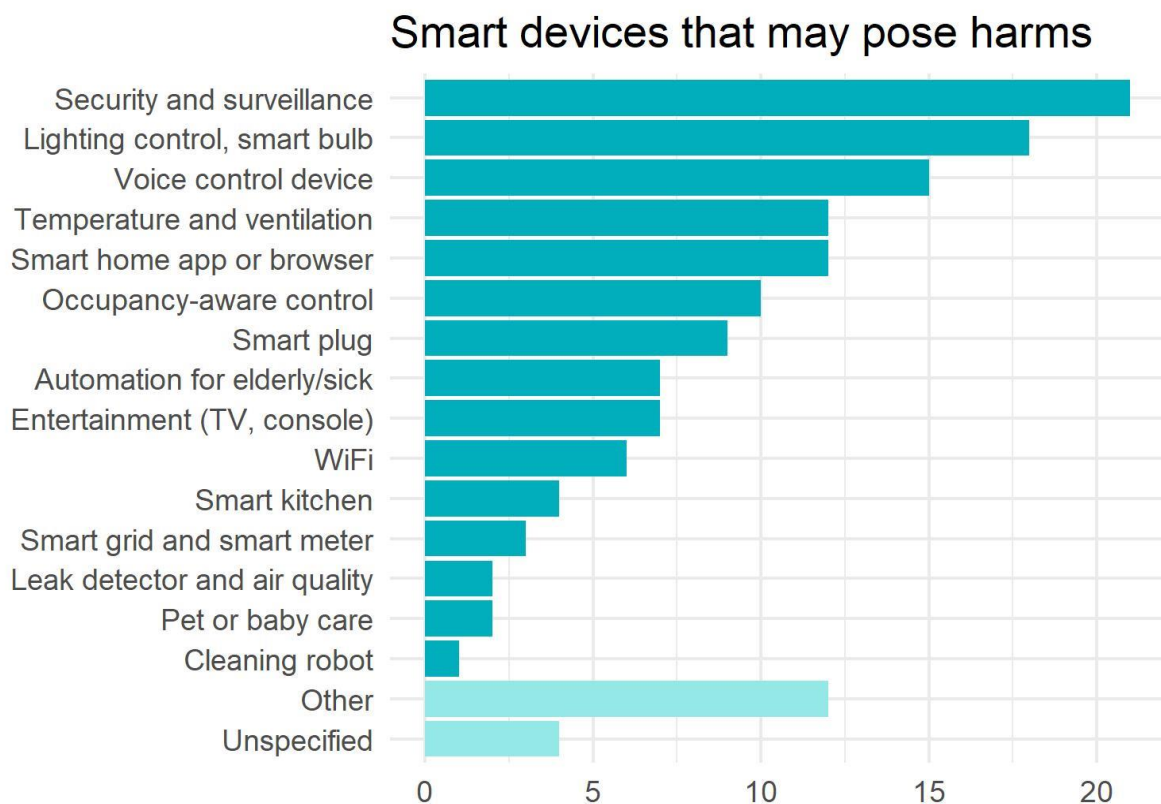

*Figure 6.* *Smart devices that may pose harms identified in systematic review*

It is important to bear in mind, however, that while the smart devices shown here may have been selected due to their actual digital harms or vulnerabilities, their selection may also be driven by the different uptake of devices in home settings, and even by researchers' preferences or the ease with which they can study some appliances. A report published by YouGov (2020) showed that the most common type of smart device in UK households are smart meters (18% at the time of the study), followed by smart speakers (11%), thermostats (6%), lighting (5%) and security systems (3%). Another survey published by techUK (2021), which did not include smart meters, showed that 58% of respondents own smart TVs, 39% smart speakers, 24% smart fitness, 15% smart thermostats, and 12% smart lighting. We thus find no direct correspondence between the most common types of devices identified in our review and the uptake of smart home devices, which shows that the usage of smart meters and smart TVs, for example, is more widespread than that of security and lighting systems. There is no data available to understand which types of devices are more commonly affected by digital harms.

Further, we recorded data about the digital harms identified for different types of smart home devices (see Table 5) and the types of data that pose digital harms in each case (Table 6). The studies included in the systematic review identified that DoS/DDoS and privacy intrusions are more common in the case of security and surveillance systems, while hacking is more commonly identified for lighting systems and voice control devices. For instance, [30] disable smart security systems through man-in-the-middle DoS attacks, [25] access private traffic data from cameras via WiFi sniffing, and [46] apply adversarial models to obtain information about household members and their routine activities from messages between smart lighting devices. DoS/DDoS attacks are also the most commonly identified type of harm in the case of WiFi and entertainment devices, while hacking is more common in the case of temperature and ventilation systems, smart home apps and browsers, smart plugs, and smart kitchen appliances. Privacy intrusions are the most common type of harm for automation of elderly/sick and smart grids and meters.

*Table 5.* Digital harms identified for each smart home device

|  | Privacy intrusion | Hacking | Malware | DoS | Stalking |
|---|---|---|---|---|---|
| Security and surveillance | 10 | 9 | 7 | 12 | 2 |
| Lighting control, smart bulb | 7 | 12 | 9 | 9 | 1 |
| Voice control device | 7 | 10 | 4 | 5 | 1 |
| Temperature and ventilation | 4 | 6 | 6 | 5 | 3 |
| Smart home app or browser | 4 | 7 | 6 | 3 | 1 |
| Occupancy-aware control | 4 | 6 | 6 | 5 | 1 |
| Smart plug | 5 | 8 | 5 | 4 | 1 |
| Automation for elderly/sick | 7 | 2 | 1 | 1 | 1 |
| Entertainment | 2 | 3 | 5 | 5 | 2 |
| WiFi | 2 | 4 | 1 | 5 | 0 |
| Smart kitchen | 0 | 4 | 3 | 1 | 0 |
| Smart grid and smart meter | 3 | 0 | 0 | 0 | 0 |
| Leak detector and air quality | 1 | 2 | 2 | 2 | 1 |
| Pet or baby care | 1 | 1 | 1 | 2 | 0 |
| Cleaning robot | 1 | 1 | 0 | 1 | 0 |

Regarding the type of data that may pose vulnerabilities, network traffic data was the most mentioned in our selection of studies (24, 38.1%), followed by energy usage data (13, 20.6%), written communications (10, 15.9%), audio (9, 14.3%), image (7, 11.1%), and video (7, 11.1%). These, nonetheless, appear to vary between devices, with network traffic data being the main type of data mentioned in the cases of security systems, lighting, temperature and ventilation, occupancy-aware control, smart plugs, automation for elderly and sick, and entertainment; audio data in the case of voice control devices; and energy usage data in the case of smart grids and readers and smart kitchens. To mention some examples, [50] analyse the eavesdropping of network traffic data from wireless cameras to identify the presence of people in the house, and [55] study malware used to access network traffic at flow-level granularity from a variety of security, lighting and occupancy-aware devices. Other types of data not covered in Table 6 included network system information, such as access points, IP addresses, and log-in credentials. We also note that, in the case of cleaning robots, no specific type of vulnerable data was identified, but [37] analysed their insecure firmware more generally.

*Table 6.* Type of data that may pose vulnerabilities for each smart home device

|  | Network traffic | Energy usage | Written comms | Audio | Image | Video |
|---|---|---|---|---|---|---|
| Security and surveillance | 12 | 3 | 5 | 1 | 3 | 3 |
| Lighting control, smart bulb | 10 | 3 | 4 | 1 | 1 | 1 |
| Voice control device | 4 | 4 | 2 | 6 | 2 | 1 |
| Temperature and ventilation | 5 | 4 | 4 | 1 | 2 | 2 |
| Smart home app or browser | 7 | 5 | 4 | 1 | 3 | 3 |
| Occupancy-aware control | 5 | 4 | 2 | 1 | 2 | 1 |
| Smart plug | 6 | 4 | 1 | 1 | 1 | 1 |
| Automation for elderly/sick | 3 | 1 | 2 | 2 | 2 | 2 |
| Entertainment | 3 | 1 | 2 | 1 | 2 | 2 |
| WiFi | 2 | 1 | 2 | 0 | 0 | 0 |
| Smart kitchen | 1 | 2 | 0 | 0 | 0 | 0 |
| Smart grid and smart meter | 0 | 3 | 0 | 0 | 0 | 0 |
| Leak detector and air quality | 1 | 1 | 2 | 1 | 1 | 1 |
| Pet or baby care | 1 | 0 | 1 | 0 | 0 | 0 |
| Cleaning robot | 0 | 0 | 0 | 0 | 0 | 0 |

## 4.4 Approaches to mitigate digital harms of smart homes

Finally, we also recorded data about the recommendations mentioned in each study to mitigate the digital harms of smart devices. 56 studies (88.9%) included explicit recommendations to mitigate digital harms. As shown in Figure 7(a), the vast majority of studies focused on technical improvements (55, 87.3%), while fewer mentioned prevention based on education (10, 15.9%) and change in policy (2, 3.2%). No study mentioned other forms of prevention, such as prevention based on control over victims or perpetrators.

Studies that focus on technical improvements, however, take highly dissimilar approaches. To mention a few examples, a variety of approaches are proposed to better identify malicious intrusions, including decision tree models [6], deep learning models that learn from time-series data [5, 26], machine learning trained from datasets of users with similar characteristics [10], automated intrusion detection systems that adapt to new threats [15, 19], and distance-based verification procedures to identify unintended pairing of IoT devices [56]. Others focus on improving the technical specification of traffic packets to better conceal their content: [59] propose making randomly occurring deceptive connections, [4, 25] replaying fake packages and flows at random times, [31] appending randomised amounts of bytes to each connection, [23] applying differential privacy to better conceal packets, and [30] adding information about the last message sent in each packet, so devices can easily identify if a device has been corrupted. Other technical recommendations include over-the-air firmware update systems to quickly address vulnerabilities of devices [52], not allowing individual devices to freely connect themselves to the network (only through a control hub) [14], and hardware and privacy moderation algorithms to protect data [22].

Several studies also mention the need to provide training and education for users, for example, in [14], "users should be encouraged to educate themselves on the aspects of cybersecurity to increase their ability to identify and respond to cybersecurity risks within smart homes" (p. 735), with a particular focus on those with cognitive impairment and deficits in [63], social workers in [34], and developers in [39]. [27] propose more stringent regulations and certification programmes.

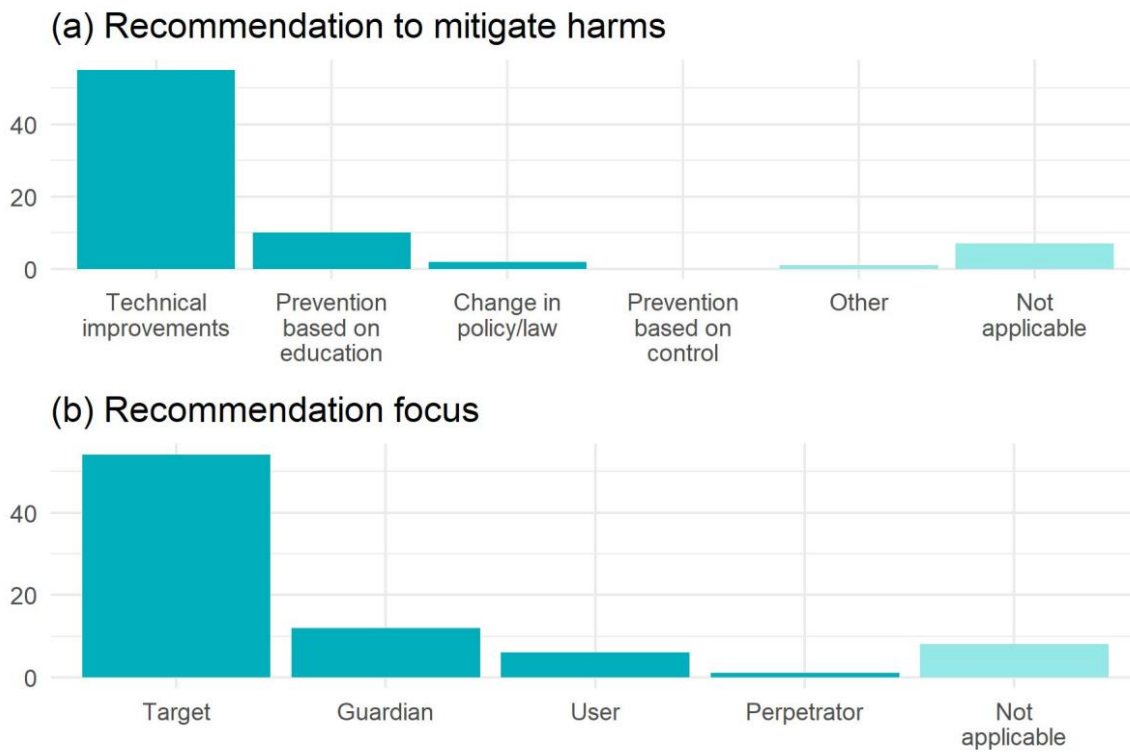

*Figure 7.* Recommendations to mitigate digital harms of smart devices

Most studies focus their recommendations on smart devices (54, 85.7%), while the proportion of studies that propose recommendations focused on the potential guardians (12, 19.0%), users (6, 9.5%) and perpetrators (1, 1.6%) is relatively small (Figure 7(b)). For instance, [34] focuses on the training of social workers to better assist victims of domestic abuse, and [28] argues developers and vendors should undertake ongoing threat assessments to improve the technical specifications of devices. Others propose more imaginative solutions, such as encouraging users to place moving objects (e.g., a clock) in front of smart cameras to continuously trigger the system and prevent offenders from detecting when users are not at home [25]. [34] propose multi-factor authentication systems prevent household members from stalking each other through companion apps.

## 5. Discussion and conclusions

The use of smart home devices is rapidly growing across the globe. Between 2017 and 2021, the use of devices such as smart thermostats, smart TVs, and smart lighting increased by over 50% in the UK, and smart doorbells and smart speakers by over 75% (techUK, 2021). Similar trends are seen in the US, where over 65% of residents own smart devices (Harvey, 2022). With the rapid uptake of smart technologies at home, it becomes vital for developers and vendors, as well as users and policymakers, to fully understand their benefits as well as their potential risks and barriers.

While there is a growing body of academic research exploring the potential harms of smart devices, there is still an overall lack of information about the nature and extent of these harms, and no public records offer insights into this. We argue that the field is now at a point where unique studies about the digital harms of smart devices can be synthesised to obtain a comprehensive overview of the digital harms of smart devices. Thus, this article has presented a first-of-its-kind systematic review of the privacy- and security-related harms of smart home technologies. Following the PRISMA protocol in two widely used academic databases, seven researchers selected a sample of 63 studies that met a set of inclusion criteria and extracted information from them. This systematic review offers an overview of smart home devices and attributes that may pose digital harms, classifies these digital harms, and summarises approaches to mitigate them.

Our review identified that the majority of existing studies focus on privacy intrusions as a prevalent form of harm against smart homes. Privacy intrusions can take the form of non-criminal (e.g., uncontrolled access by medical practitioners and carers to private data recorded in care monitoring devices; Alkhatib et al., 2019) as well as criminal behaviour (e.g., when information obtained from smart homes is subsequently used to control household members or target houses for burglary; Hodges, 2021), which in turn affects the types of actions companies and law enforcement should put in place to prevent and respond to them. Other factors that influence the private and public responses to digital harms include the type of device and data linked to each harm, and the nature of the harm itself. This is the reason why in this study we recorded information not only about the most prevalent harm types, but also classified these harms and the variety of harm-affected devices.

Other types of harms that are less common in the literature include hacking, malware and DoS/DDoS attacks targeting a variety of IoT home devices. Fewer articles studied stalking incidents, and one of them, Sikder et al. (2020), found that the way in which access control settings in smart homes are designed leads to users being granted undesired full access control to all smart home ecosystems (e.g., AirBnB guests). While the differences in the prevalence of the types of harms identified in this systematic review may indeed reflect real-world patterns, the frequencies observed here are likely affected by the overall interests of researchers and funders, and that it is easier or more convenient to study some harm types over others. Thus, while this systematic review provides valuable information about the types of harms that researchers have so far identified, it is necessary for researchers and public authorities to collaboratively work on new ways to more accurately estimate the extent and nature of digital harms. Some consider that creating public-private partnerships for data sharing and evidence-based prevention in the context of smart homes is essential to further understand their benefits and harms, and in turn put measures in place for prevention (Buil-Gil et al., 2022).

Moreover, in order to better understand these harms and derive effective prevention mechanisms, either technical, social, or socio-technical, we argue it is important to classify them according to their nature, methods and objectives (Lin and Bergmann, 2016; McGuire and Dowling, 2013; Wall, 2001). We found that harms identified in extant academic literature tend to cluster on incidents at the intersection of cyber-dependent, access and trespass, and cyber-dependent, confidentiality and trespass. We have seen several examples of harms with these characteristics, as presented in the academic literature as well as in known cases, but we have also seen examples of other types of harms that are either less commonly identified or fully neglected in existent research. While our systematic review is important to gain a better understanding of the nature of the harms of smart devices, it also identifies gaps in research that should be better addressed in the future. For instance, we found research gaps regarding harms at the intersection of cyber-enabled and deception, and cyber-enabled and porn, which nonetheless do exist in the real-world (e.g., data retrieved from smart homes being used for identity theft or to assist fraud, disseminating sexually explicit images of children obtained from monitoring devices).

Another key finding of this systematic review is that digital harms, and data associated with these harms, may vary extensively across smart home devices. For instance, according to data extracted from this review, while harms associated with security and surveillance systems have been mainly linked to DoS attacks and privacy intrusions arising from insufficient protection of network traffic data, voice control devices (e.g., Google Home, Amazon Alexa) are more commonly associated with the hacking of audio data. And while lighting control systems are commonly linked with the hacking of network traffic data, smart grids/meters are mainly linked to privacy intrusions of energy usage. This type of information may indeed be essential to propose and design better prevention mechanisms that adapt to the types of data vulnerabilities and harms of each specific device, user, and context. In a similar vein, these findings can help inform policy and legislation such as the UK PSTI Bill.

The vast majority of studies included in our systematic review propose explicit measures to mitigate the digital harms identified. Most of these recommendations focus on technical improvements with different aims, mostly related to improving intrusion detection systems, data protection and concealment mechanisms, and software updates. Fewer mentioned hardware improvements. While we found a considerable proportion of studies proposing technical recommendations for harm prevention, very few articles mentioned social prevention mechanisms such as improving the education of users or developers, and only two articles described the need to apply changes in law and policy. Relatedly, most of these recommendations focused on the target, with fewer considering harm reduction and prevention from the perspective of the guardian, user, or perpetrator (Leukfeldt and Yar, 2016). This article thus identifies another important gap in research: the need to consider and evaluate the effectiveness of social and socio-technical prevention approaches that focus on the guardian, user, and perpetrator.

Some of these gaps in research could, and perhaps should, be addressed through cross-disciplinary initiatives involving researchers from different fields. We have observed an overall lack of multidisciplinary work in this domain, with not a single study involving researchers from across both technical and health or social sciences disciplines. For instance, our review found evidence that while crossing physical and political boundaries does not appear to be an issue for collaborative work (i.e., 13 studies involved authors from multiple countries), crossing disciplinary boundaries appears much more challenging for researchers interested in the study of smart homes. This is likely to be the primary driver for some of the research gaps identified, including the lack of research about cyber-enabled harms, and incidents related to deception, violence, and porn, and the main focus on solely technical prevention mechanisms to improve the protection of smart devices. We argue enhancing cross-disciplinary work in this domain is not only important to better address the wider variety of harms that affect devices, and the wider possibilities of harm reduction strategies, but to better research them. While most studies in our systematic review were described as experimental or quasi-experimental by design, few of them consider the selection of randomised control and trial groups, which is considered a fundamental requirement for experimental designs in many disciplines. Similarly, very few studies in our sample apply mixed-methods (i.e., combining quantitative and qualitative data analysis) to better understand the harms of smart homes, and not a single study applies meta-analytical designs to compare findings presented in multiple studies. Few studies considered the experiences of victims in the assessment and response to the security and privacy threats of smart homes (Leitão, 2019). As has been noted regarding the study of wearable technology (Ferreira et al., 2021), research in the field of smart homes will undoubtedly benefit from further enhancing principles of cross-disciplinarity and considering the voices of everyone involved in the design, development, and use of smart home technologies and devices.

techUK (2021). *The state of the connected home 2021: A year like no other*. Available from: https://spark.adobe.com/page/LCRPh1X14fjDM/ (Accessed 11 August 2021).

Teng, C.C., Gong, J.W., Wang, Y.S., Chuang, C.P., and Chen, M.C. (2017). Firmware over the air for home cybersecurity in the Internet of Things. In *19th Asia-Pacific Network Operations and Management Symposium* (pp. 123-128). IEEE.

Tran, B., Pan, S., Liang, X., and Zhang, H. (2021). Exploiting physical presence sensing to secure voice assistant systems. In *IEEE International Conference on Communications* (pp. 1-6). IEEE.

Tushir, B., Dalal, Y., Dezfouli, B., and Liu, Y. (2021). A quantitative study of DDoS and E-DDoS attacks on WiFi smart home devices. *IEEE Internet of Things Journal*, 8(8), 6282-6292.

Tzezana, R. (2016). Scenarios for crime and terrorist attacks using the internet of things. *European Journal of Future Research*, 4, 18.

Ullrich, F., Classen, J., Eger, J., and Hollick, M. (2019). Vacuums in the cloud: Analyzing security in a hardened IoT ecosystem. In *13th USENIX Workshop on Offensive Technologies*. USENIX.

US Department of Justice (2017). *Justice Department Announces Charges and Guilty Pleas in Three Computer Crime Cases Involving Significant Cyber Attacks*. U.S. Attorney's Office. Available from: https://www.justice.gov/usao-nj/pr/justice-department-announces-charges-and-guilty-pleas-three-computer-crime-cases (Accessed 2 August 2022).

Vidal-González, S., García-Rodríguez, I., Aláiz-Moretón, H., Benavides-Cuéllar, C., Benítez-Andrades, J.A., García-Ordás, M.T., and Novais, P. (2020). Analyzing IoT-based botnet malware activity with distributed low interaction honeypots. In A. Rocha, H. Adeli, L.P. Reis, S. Costanzo, I. Orovic and F. Moreira (Eds.), *Trends and innovations in information systems and technologies*, Volume 2 (pp. 329-338). Cham: Springer.

Wall, D. (2001). *Crime and the Internet*. New York: Routledge.

Wan, Y., Xu, K., Wang, F., and Xue, G. (2021). IoTAthena: Unveiling IoT device activities from network traffic. *IEEE Transactions on Wireless Communications*, 21(1), 651-664.

Wang, C., Kennedy, S., Li, H., Hudson, K., Atluri, G., Wei, X., Sun, W., Wang, B. (2020). Fingerprinting encrypted voice traffic on smart speakers with deep learning. In *WiSec '20: Proceedings of the 13th ACM Conference on Security and Privacy in Wireless and Mobile Networks* (pp. 254-265). New York: ACM.

Wang, Q., Ji, S., Tian, Y., Zhang, X., Zhao, B., Kan, Y., Lin, Z., Lin, C., Deng, S., Liu, A. X., and Beyah, R. (2021). {MPInspector}: A systematic and automatic approach for evaluating the security of {IoT} messaging protocols. In *30th USENIX Security Symposium* (pp. 4205-4222). USENIX.

Wang, Y., Li, X., Jia, P., Yang, Y., and Wang, H. (2021). Sensitive instruction detection based on the context of IoT sensors. In *51st Annual IEEE/IFIP International Conference on Dependable Systems and Networks Workshops* (pp. 121-128). IEEE.

Weber, R.H. (2010). Internet of Things – New security and privacy challenges. *Computer Law & Security Review*, 26(1), 23-30.
30